\documentstyle[12pt,graphicx]{article}

\begin{document}

\title{Calculation of the Condensate Fraction in~Liquid Helium-4}

\author{
A.~A.~Rovenchak\footnote{andrij@ktf.franko.lviv.ua}
\ and
I.~O.~Vakarchuk\footnote{vakarch@ktf.franko.lviv.ua}\\
Department for Theoretical Physics, \\
Ivan Franko Lviv National University,\\
12 Draghomanov Str., Lviv, Ukraine, UA--79005}

\date{}
\maketitle

\begin{abstract}
We adduce the results of the condensate fraction calculation for liquid
helium-4. The method is derived from the first principles and involves
minimum assumptions. The only experimental quantity we need in our
calculations is the static structure factor which is easily measurable.
We use the approximation in which expressions contain one
summation in the wave vector space (or one integration it the radius
vector space) in order to demonstrate the validity of our method.
The computed
values of the condensate fraction lie within $8.8\div14\%$
depending on the model potential for the short-range interaction.
This result agrees with recent experimental measurements and
numerical estimations.


\end{abstract}

\pagebreak

\section{Introduction}
The phenomenon of Bose--Einstein condensation and superfluidity in Bose-systems
is a subject of great interest.
Superfluidity in liquid helium-4 can be easily observed now but
the measurements on the condensate fraction remain a difficult
problem because of the strong interactions in this fluid.
Researchers, both experimentalists and theorists, now generally concur that
$\sim10\%$ of atoms are in the lowest (zero momentum) state at
zero temperature. The relative number of atoms with zero momentum
is called condensate fraction, here we denote it as $f$.

The problem of the calculation of condensate fraction in liquid helium-4 was
first considered in the classical paper by Penrose and Onsager~\cite{PenroseOnsager}.
The authors used a very rough approximation of hard spheres for the ground-state wavefunction and
calculated this quantity as approximately 0.08 or 8\%.

During recent twenty years different methods for measurements and
theoretical calculations of the condensate fractions were used.
Neutron scattering is used to draw the information on the
momentum distribution and hence on the condensate fraction.
Sears {\it et al}~\cite{SearsSvensson82} obtained the value
13.9\%. Later, similar value of 13.3\% was obtained by
Sears~\cite{Sears83} using the data on the temperature variations of the average
kinetic energy.

Numerical study based on the variational method were carried out
in the series of works by Manousakis, Pandharipande {\it et
al}~\cite{Manousakis85,Manousakis85a,Manousakis91} by means of
Jastrow wavefunction with three-body corrections. Their results
vary within $8.2\div10.3\%$.

The numerical estimations of the condensate fractions were made
also in the series of works by
Vakarchuk {\it et al} and
Vakarchuk~\cite{Vakarchuk87,Vakarchuk90UFZh,Vakarchuk90TMF} leading
to the results of $3.7\div8.8\%$. The methods used in these papers
are based on the mean-spherical approximation for the structure
functions of liquid helium-4 at zero
temperature~\cite{Vakarchuk87}, direct quantum-mechanical computations from the first
principles~\cite{Vakarchuk90UFZh,Vakarchuk90TMF}.

The value of approximately 10\%  was obtained for $f$ by Sokol and
collaborators by means of deep-inelastic neutron scattering at
high-momentum transfer~\cite{Sokol89,Sokol90,Sokol91}.

Monte Carlo (MC) methods were also used to study the question of the
condensate fraction. The result of Whitlock and Panoff~\cite{Whitlock87}
is $\simeq9\%$ (Green Function MC).
Ceperley and Pollock~\cite{Ceperley87} obtained value $\simeq7\%$
using Path Integral MC technics at 1.18~K.
Also, Moroni {\it et al}~\cite{Moroni97} gives 7.2\%
from Diffusion MC simulations.

Approximately in the same time the experimental result of
$\simeq6\%$ was obtained by Azuah {\it et al}~\cite{Azuah97}.

Series of papers by Mayers {\it et al}~\cite{Mayers96,Mayers97}
was devoted to both calculation of the condensate fraction based on
phenomenological assumptions \cite{Mayers96} (result is 9.9\%) and
measurements on the high-energy scattering \cite{Mayers97} ($f=15\pm4\%$
at 1.3~K).

Recent high precision measurements of the dynamic structure factor
allowed Glyde {\it et al}~\cite{Glyde00}
to derive the information on the condensate
fraction at zero temperature and its temperature dependence. They report
$f=7.25\pm0.75\%$ at 0~K.

A semi-phenomenological method for extracting the condensate fraction data
was proposed recently by Rinat and Taragin~\cite{RinatTaragin01}.
Authors obtained $f$ at different temperatures and extrapolated
their results to 0~K as $f=9.0\pm0.3\%$.

The aim of this work is to show a possibility of calculation of
the condensate fraction without drawing any special huge
computational efforts.

The method we use was worked out in~\cite{Vakarchuk90TMF}. The
condensate fraction is calculated directly from the single-particle
distribution function $F_1(R)$ as its long-range limit:
$$
f=N_0/N=\lim_{R\to\infty} F_1(R)
$$
where $N_0$ is the number of particles in zero momentum state and
$N$ is the total number of particles in the system.

Our technics do not involve any phenomenological assumptions. For
the computations we use as input only one experimental
quantity, the static structure factor of helium at zero temperature.
It is the advantage comparing to those
methods for which such quantities as the dynamic structure
factor is necessary. Its precise measurement is much more complicated
problem comparing with that of the static structure factor.

The method is fully controlled, i.~e., we
always know which effects are neglected, and it is not difficult
to take them into account in the next approximation (in
principle). Here we present only the first approximation (RPA) but
the expressions are easily extendable in order to take subtler
effects like many-particle interactions.

The only key problem arising while extending our method further is
the knowledge of the static
three-particle structure factor. Unlike usual structure factor, it is not
an easily measurable quantity already. One of the ways for its calculation
we see in such a scheme: expressions for the three-particle distribution
function $F_3$ (beyond superposition approximation) containing pair distribution function $\to$
6-dimensional Fourier transformation of $F_3$ leading to the three-particle
structure factor. A separate paper will be devoted to this problem.

We consider three different approaches for the short-range repulsive
interactions.
In the first one, they are not taken it into account explicitly at all.
In the second approach this repulsion is modeled by
Meyer's function $e^{(A/R)^n}-1$, $R$ is the radius-vector absolute value.
We chose $A=2.1$~\AA{} and $n=12$. This potential is referred here as
`almost hard spheres' (AHS)~\cite{Ours2001}.
The third model is the potential of hard spheres (HS) with diameter 2.1~\AA.

\section{Calculating procedure}
Principal calculating procedure consists of the following steps:

\begin{enumerate}
\item {\bf Input}: Experimental data on the structure factor
\cite{SvenssonSears80}
\label{ProcInput}
(should be converted into the zero temperature \cite{Ours2000});
\item {\bf Input}: Model for the short-range interactions
\label{ProcModel}
(HS, AHS or none)
\cite{Ours2001};
\item {\bf Output}: Structure factor of the model system corresponding
to the short-range repulsion;
\label{Proc_Sqsr}
\item {\bf Output}: Fourier transformation of the interatomic potential
\label{Proc_nuq}
(effective) \cite{Ours2000,Ours2001};
\item {\bf Output}: Pair distribution function $F_2(R)$
\label{Proc_F2}
\cite{SvenssonSears80};
\item {\bf Result}: Condensate fraction. \label{Proc5}
\end{enumerate}

In this work, we propose a simple method for the calculation of the condensed fraction
using formulae obtained earlier in \cite{Vakarchuk90UFZh}:
\begin{eqnarray} \label{Initial}
&&f=\exp\left(I_{1A}+\Delta_1J+\ldots\right),\\ \nonumber
&&I_{1A}=-{1\over4N}\sum_{{\bf q}\neq0}(\alpha_q-1)^2/\alpha_q,\\
\nonumber
&&\Delta_1J=\varrho \int d{\bf R} \left[2 h^*(R)-h(R)+h^2(R)/4 \right],\nonumber\\
&&h(R)=F_2(R)-1, \qquad h^*(R)=\sqrt{F_2(R)}-1,\nonumber
\end{eqnarray}
dots denote terms having more than one integration over wave
vector or coordinate, $F_2(R)$ is the pair distribution function (PDF),
$N$ is the number of atoms,
and
\begin{equation}
\alpha_q=\sqrt{1+2\varrho\nu_q\biggl/{\hbar^2 q^2\over2m}},
\end{equation}
$\varrho$ is helium density, $\varrho=0.02185$~\AA$^{-3}$, $m$ is
the mass of helium atom, $m=4.0026$~a.~m.~u., $q$ is the wave
vector.

The expression for PDF within the accepted approach was obtained
in~\cite{Vakarchuk90UFZh}:
\begin{equation} \label{F2def}
F_2(R)=\exp\left\{
  {1\over N}\sum_{{\bf q}\neq0}{2a_2({\bf q})\over1-2a_2({\bf q})}
  e^{i{\bf qR}}+\dots
\right\},
\end{equation}
while the self-consistent description of short- and long-range correlations
in liquid helium-4 results in such a result for the function
$a_2$~\cite{Vakarchuk79}:
\begin{equation} \label{a2def}
a_2({\bf q})={1\over2}\left( {1\over S^{\rm s.r.}_q} - {1\over S_q} \right)
\end{equation}
with $S^{\rm s.r.}_q$ being the structure factor of the model system
(it equals 1 if the short-range repulsion is not given explicitly),
superscript `s.~r.' corresponds
to `short-range'. $S_q$ is the experimental structure factor (at 0~K).
These expressions
contain $\nu_q$ being the Fourier image of the interatomic potential in helium that has an effective nature.
It is obtained from the first principles using the collective
variables formalism in the Schr\"odinger equation, as described in \cite{Ours2000}.

On the step (\ref{Proc_nuq}) of the calculation procedure we
encountered problems while considering the HS short-range
potential. The reason is clear: a weak damping of the structure
functions at large values of wave vector or radius vector requires much more
careful numerical computations comparing to those
in two other cases.

\begin{figure}[t]
\includegraphics[width=75mm,clip]{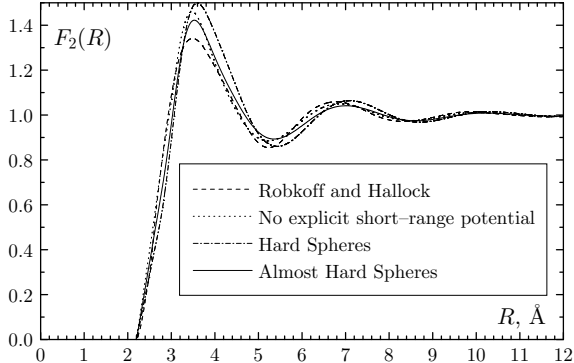}
\caption{Pair distribution function.
The dashed line corresponds to the experimental data
by Robkoff and Hallock~\cite{Robkoff81} at 1.38~K.
The dotted line represents the smoothed data in the case when no explicit
short-range repulsion is given.
The dashed-dotted and solid lines are PDFs for HS and AHS short-range
potentials respectively.}
\label{F2fig}
\end{figure}

Some usual problems appear on the step (\ref{Proc_F2}) of the
calculating procedure where we smoothed $F_2(R)$ calculated from
(\ref{F2def}) and truncated it at the distances $R<2.2$~\AA.
We put $F_2(R<2.2\ {\rm\AA})=0$ in order
to obtain a correct PDF at small
distances~\cite{Robkoff81}. We present the calculated PDFs
in comparison with the experimental (being more precise, derived from
the measurements on the liquid structure factor) one~\cite{Robkoff81}
in the Fig.~\ref{F2fig}.

\bigskip

\section{Results and discussion}
We have calculated the condensate fraction at zero temperature
using different models for the short-range repulsive part of the
potential.

The calculated values of $f$ are:
\begin{itemize}
\item No explicit short-range potential $f=14\%$;
\item AHS short-range potential $f=11\%$;
\item HS short-range potential $f=8.8\%$.
\end{itemize}
\bigskip

We expect the real value being about $10\div30\%$ less than the
calculated one due to the contribution of the higher-order terms, as
described in~\cite{Vakarchuk90UFZh}. But in the first approximation we
consider it to be a good accuracy for $f$ in liquid helium-4.

Our result for the condensate fraction is within the range of theoretical,
numerical and experimental estimations mentioned here in the
introductory part of the paper. Since our expressions are
derived from the first principles, one can use the number for $f$ calculated
in this way as a test for the condensate fraction information
indirectly extracted from the experimental measurements. This becomes
possible after the next one and two approximations are computed.

We found that the short-range repulsive part of the potential has
a significant influence on the results for the condensate fraction.
Namely, when the short-range interactions are included into the
consideration explicitly, the results appear to be essentially dependent
on the ``hardness'' of the core. A proper model for this interaction
will be found when more terms are taken into account in the expressions
(\ref{Initial}), (\ref{F2def}). The reason is that in principle the summation
of the whole series corresponds to the correct calculation of the short-range
interactions~\cite{Vakarchuk79}. On the other hand, the condensate fraction in liquid
helium-4 is a very sensitive (with respect to the order of approximation) quantity.
Thus, we can expect that having found such a model we obtain a possibility
to use simple RPA-like expressions instead of higher-order ones. By now,
the AHS ($n=12$, $A=2.1$~\AA) potential seems to be the most suitable
for this purpose.

\end{document}